\documentclass[conference]{IEEEtran}
\usepackage{amsmath}
\usepackage{array}
\usepackage{amssymb}
\usepackage{enumitem}
\usepackage{textcomp}
\usepackage{gensymb}
\usepackage{siunitx}
\usepackage{rotating}
\usepackage{xspace}
\usepackage{blkarray}
\usepackage{multirow}
\usepackage{graphicx}
\usepackage{epstopdf}
\usepackage{wrapfig}
\usepackage{xspace}
\usepackage[utf8]{inputenc}
\usepackage{mathtools}
\usepackage{mathptmx}
\usepackage{lmodern}
\usepackage{supertabular}
\usepackage{adjustbox}
\usepackage{booktabs}
\usepackage{tabularx}
\usepackage[table,xcdraw]{xcolor}
\usepackage{makecell}
\usepackage{etoolbox}

\begin{document}
%
\title{Practical Prediction of Human Movements Across Device Types and Spatiotemporal Granularities}

\author{
\IEEEauthorblockN{~Babak Alipour\IEEEauthorrefmark{1}\\ \tt\footnotesize babak.ap@ufl.edu} \and 
\IEEEauthorblockN{Leonardo Tonetto\IEEEauthorrefmark{2}\\ \tt\footnotesize tonetto@in.tum.de} \and
\IEEEauthorblockN{Roozbeh Ketabi\IEEEauthorrefmark{1}\\ \tt\footnotesize roozbeh@ufl.edu}
\and 
\IEEEauthorblockN{Aaron Yi Ding\IEEEauthorrefmark{3}\\ \tt\footnotesize aaron.ding@tudelft.nl}
\and
\IEEEauthorblockN{Jörg Ott\IEEEauthorrefmark{2}\\ \tt\footnotesize ott@in.tum.de} \and
\IEEEauthorblockN{Ahmed Helmy\IEEEauthorrefmark{1}\\ \tt\footnotesize helmy@ufl.edu}
\and
\IEEEauthorblockA{
\IEEEauthorrefmark{1}
University of Florida, USA\\
} \and
\IEEEauthorblockA{
\IEEEauthorrefmark{2}
Technical University of Munich, Germany\\
} \and
\IEEEauthorblockA{
\IEEEauthorrefmark{3}
TU Delft, The Netherlands\\
}
}

\makeatletter
\patchcmd{\@maketitle}
  {\addvspace{0.5\baselineskip}\egroup}
  {\addvspace{-1\baselineskip}\egroup}
  {}
  {}
\makeatother

\maketitle
\thispagestyle{plain}
\pagestyle{plain}

\begin{abstract}
Understanding and predicting mobility are essential for the design and evaluation of future mobile edge caching and networking.
Consequently, research on prediction of human mobility has drawn significant attention in the last decade.
Employing information-theoretic concepts
and machine learning methods, earlier research has shown evidence that human behavior can be highly predictable. 

Despite existing studies, more investigations are needed to capture intrinsic mobility characteristics constraining predictability, and to explore more dimensions (e.g. device types) and spatio-temporal granularities, especially with the change in human behavior and technology. We analyze extensive longitudinal datasets with fine spatial granularity (AP level) covering 16 months.
The study reveals \textit{device type} as an important factor affecting predictability. Ultra-portable devices such as smartphones have ”on-the-go” mode of usage (and hence dubbed ”\textit{Flutes}”), whereas laptops are  ”sit-to-use” (dubbed ”\textit{Cellos}”). 

The goal of this study is to investigate practical prediction mechanisms to quantify predictability as an aspect of human mobility modeling, across time, space and \textit{device types}. We apply our systematic analysis to wireless traces from a large university campus. We compare several algorithms using varying degrees of temporal and spatial granularity for the two modes of devices; \textit{Flutes} vs. \textit{Cellos}. Through our analysis, we quantify how the mobility of \textit{Flutes} is less predictable than the mobility of \textit{Cellos}. 
In addition, this pattern is consistent across  various  spatio-temporal  granularities,  and  for
different methods (Markov chains, neural networks/deep learning, entropy-based estimators).
This work substantiates the importance of predictability as an
essential aspect of human mobility, with direct application in predictive caching, user behavior modeling and mobility simulations.  

\end{abstract}

%
\IEEEpeerreviewmaketitle

\section{Introduction \& Related work}

In recent years, large-scale research on human mobility has thrived due to the availability of location data collected from portable computing and communication devices, such as laptops, smartphones, smartwatches and fitness trackers. 
One particular aspect of human mobility that has gained a lot of attention lately is predictability. 
Prediction techniques constitute fundamental mechanistic building blocks for many mobile protocols and applications, ranging from resource allocation to caching and recommender systems, among others \cite{Siris2016,Lathia2015anatomy}.

The seminal work by \cite{Song2010limits}, utilizing cellular network data, established an approach towards understanding and measuring predictability of human mobility patterns, with their equally important contribution with respect to the data-driven analysis of large mobile populations, and their efforts in devising a framework to study the theoretical limits of predictability.
The methods introduced in their framework are founded in information theory and have since been extensively applied in the area of mobility modeling and prediction.
Later studies that built on \cite{Song2010limits} addressed either the specifics of the prediction problem (e.g., different formulations \cite{Smith2014} of the individual's change of location, analyzed different contexts of mobility) or the shortcomings of the original approach (that relied on coarse spatio-temporal granularity).
Authors in \cite{Cao2017infocom} used Wireless LAN (WLAN) traces from a university campus network and reported multi-modal entropy distributions which can be partially explained by the demographics of the population (\textit{i.e.}, age, gender, major of studies). Other entropy based studies include vehicular mobility \cite{Li2014its,Wang2015,Gallotti2013}, online social behavior \cite{takaguchi2011predictability,sinatra2014entropy}, complex systems \cite{Hanel2011}, cellular network traffic \cite{Zhou2012} and public transport utilization \cite{Goulet-Langlois2017}.
In addition, devices' form factor affects the mode of usage and varied traffic profiles
(\cite{maier2010first, chen2012network, Kumar2013, Alipour2018}), but these studies either do not consider predictability or do not account for different spatio-temporal resolutions.
We have chosen our methods based on the literature to measure and compare both theoretical and practical limits of predictability for \textit{Flutes} and \textit{Cellos}, with varying degrees of spatio-temporal granularity, while also looking at the correlation of prediction accuracy with mobility and network traffic profiles using extensive fine-granularity traces (based on our earlier work in \cite{Alipour2018}).

The \textit{main} questions addressed in this study are:
i. How different are \textit{Flutes} and \textit{Cellos} in terms of predictability? 
ii. How does the predictability of these device types change with different \textit{spatio-temporal granularity} (5, 15, 30 min, 1 hour and 2 hours; access point and building level)?
iii. Does the \textit{choice of method} or predictor (\textit{e.g.} Markov Chain, neural networks such as LSTM and CNN, BWT or LZ based estimators, which are introduced in Section \ref{sec:approach}) significantly alter the answers to aforementioned questions?

This study provides the following main contributions:
1. Quantifying the differences of \textit{Flutes} and \textit{Cellos} for prediction analysis, evaluated on a real-world large-scale dataset.
2. Comparison of several well-known algorithms (Markov Chains, Neural Networks) and LZ/BWT-based theoretical bounds across different time and space scales for Flutes and Cellos.
3. Use of prediction accuracy as part of the user profile for modeling, and investigation of its correlation with a combination of network traffic and mobility features.

The paper is structured as follows: First, the main approach and methods are presented in Sec. \ref{sec:approach}. Then, the details of the dataset and experiment setup are discussed in Sec. \ref{sec:datasets}. The experiment results are presented in Sec. \ref{sec:exp}. Sections \ref{sec:discussion} and \ref{sec:conclusion} present the discussion on potentials implications of the findings and conclude the paper.

\section{Main Approach \& Methods}
\label{sec:approach} 

We investigate two methods to measure predictability; a theoretical method based on entropy, and a systems method based on practical predictor algorithms. Following we provide the entropy estimation based definition and discuss the different algorithms studied in this paper, including a reference-point Markov Chains approach, and a more sophisticated deep learning approach.

\subsection{Entropy Estimation}
\label{subsec:entropyestimators}
\textit{Entropy} is defined as the level of order (or disorder) of a system, and is founded on information theory. It has been adopted in previous studies to establish bounds on predictability under certain assumptions \cite{Song2010limits, Smith2014}. 
We utilize it in our study to gauge the performance of our practical predictors. For a random process, this metric is sensitive to both the relative frequency of events and their inter-dependencies \cite{Goulet-Langlois2017}. 
To estimate a baseline of predictability, we compute the \textit{time-uncorrelated} entropy ($\mathcal{S}^{\text{unc}}$) which only takes into account the frequency of the observed events. For the upper-bound of predictability we compute two \textit{time-correlated} estimators based on compression algorithms ($\mathcal{S}^{\text{lz}}$ and $\mathcal{S}^{\text{bwt}}$) which also consider the memory of the system.
We define \textit{maximum predictability} as the probability of predicting the most likely state of $x_i$ given a state $x_j$, which is computed from the entropy $S$ of a given sequence of events based on \cite{Song2010limits}, with the refinements proposed by \cite{Smith2014}.
For a complete description on \textit{entropy estimation}, we kindly refer the reader to \cite{Gao2008} and \cite{Cai2004}.

\subsection{Predictors}
\label{predictors}

\subsubsection*{Markov Chain-based predictor}
\label{sec:markov}
A Markov chain (MC) with a discrete state space has been applied for user mobility prediction \cite{Song2004infocom-predictors,lu2013approaching}. 
In an order-\(k\) Markov predictor, the state space consists of tuples of $k$ location names (e.g., AP), where the next location prediction depends solely on the most recent preceding \(k\)-tuple.
We build the model on the data so that observed k-tuples comprise the states. The transition probabilities are learned based on the frequency of appearances of such a transition in observations. The probability for a transition from the current state $S = X_iX_{i+1}...X_j$ to $X_{i+1}X_{i+2}...X_jX_{j+1}$ where $j-i=k$ and each $X_i$ is the symbol for each location, is represented as $P(X_{j+1} = c\ |\ S = X_iX_{i+1}...X_j)$ for all $c$ observed in data and is learned based on the reappearance frequency of such a sequence.
If the predictor of order $k$ encounters a new sequence that has never seen before, it falls back to the lower, $k-1$ order recursively. 
The base case is O(0) which is simply the frequency distribution of all symbols observed so far.

\subsubsection*{Deep learning}
\label{sec:deeplearning}
Recent approaches to sequence prediction use deep Recurrent Neural Networks (RNN) or Convolutional Neural Networks (CNN).
Recurrent neural networks have loops within their cells, allowing information to persist and thus enabling the neural network to connect previous information to make a reasonable prediction of the future. Certain types of RNNs are capable of learning long-term dependencies.
There are multiple variants of RNNs, including Long short-term memory (LSTM) \cite{lstm} and Gated Recurrent Unit (GRU) \cite{gru}.
These networks can learn dynamic temporal patterns and have successfully been applied
in speech recognition, text-to-speech engines and predicting next location \cite{DLsurvey, karatzoglou2018seq2seq}.  

CNNs learn convolutional filters to extract latent information across the data (i.e. 1D CNNs learn different temporal locality patterns) and use that information for predicting the next location.
In our study, we use a multi-layer LSTM and 1D CNN to predict movements of users based on similar input tuples used for MC-based predictors.
Neural networks are computationally expensive and require hyper-parameter tuning.
Thus the deep model is run only on a sample of users in this study.
One goal of this study is to analyze the payoff (and cost) of adding complexity to the predictor (e.g. LSTMs), versus the simpler MC-based predictors.

\section{Datasets \& Experimental Setup}
\label{sec:datasets}
To study the regularity of human behavior, we performed a data-driven analysis applying our methods to a university campus WiFi traces from the University of Florida. 
The datasets were collected from networks providing wireless access to a large number of portable devices via access points deployed in non-residential areas, including classrooms, computer laboratories, libraries, offices, administrative premises, cafeterias, and restaurants.

Every trace entry contains a unique user identifier (\textit{uuid}), time-stamp and an access point unique identifier (\textit{apid}). Based on the \textit{apid}'s string we are able to identify the building as well as the room in which an access point (AP) was located. Only the geographical coordinates of buildings are known. Table \ref{tab:dev_stats} contains a brief summary of the dataset with mean ($\mu$) and standard deviation (\textit{std}), where $N_{\text{ap}}$ is number of unique access points observed per device, $N_{\text{day}}$ number of unique days with at least one record, $N_{\text{rec}}$ number of records during data collection, and \textit{total} number of devices available for at least 7 days and accessed more than 5 APs.\footnote{Transient devices are not counted to ensure the analysis is carried out on devices that are mobile and benefit from predictive systems the most, while stationary devices (e.g. plugged-in Cellos) and guests that never return to campus are ignored.}

\begin{table*}[!ht]
\centering
\caption{AP logs sample data columns}
\begin{adjustbox}{max width=\textwidth}
  \begin{tabular}{*{6}{|c}|}
\hline
  User IP & UUID & AP name & AP MAC & Lease begin time & Lease end time  \\
\hline
10.130.90.3 &00:11:22:00:00:00& b422r143-win-1& 00:1d:e5:8f:1b:30 &1333238737&  1333238741\\
\hline
\end{tabular}
\end{adjustbox}
  \label{tab:dhcp_sample}
  \vspace*{-0.3cm}
\end{table*}

\subsection{UF traces}
\label{subsec:uf-trace}

The UF traces were collected for 16 months (September/2011-December/2012) and contain over 1700 wireless access points (APs) deployed in 140 buildings which were used by 300K devices. 
A sample (sythentic) record is shown in Table \ref{tab:dhcp_sample}.
Its raw records were captured from associations and sessions timeout in which the unique user id (\textit{uuid}) was the MAC address. These \textit{uuid} although hashed, still contained the Organizationally Unique Identifier (OUI)\footnote{http://standards.ieee.org/faqs/regauth.html\#17} allowing us to distinguish \textit{Flutes} and \textit{Cellos}, as detailed in \cite{Alipour2018}. 
All collected WiFi traces are processed as discrete time-series, defined next.


\begin{table}[!ht]
\centering
\caption{Statistics per device available for at least 7 days \& accessed more than 5 APs.}
\label{tab:dev_stats}
\begin{tabular}{@{}lccccccc@{}}
\toprule
    & \multicolumn{2}{c}{$N_{\text{ap}}$} & \multicolumn{2}{c}{$N_{\text{day}}$} & \multicolumn{2}{c}{$N_{\text{rec}}$} & \multirow{2}{*}{Total Devices} \\
    & $\mu$        & std        & $\mu$         & std        & $\mu$        & std        &                     \\ \midrule
UF  & 127.3        & 142.3      & 63.5          & 59.2       & 1861         & 5121       & 138028 \\ \bottomrule
\end{tabular}
\vspace*{-0.2cm}
\end{table}

\subsection{Discrete-time Series}
\label{subsec:discretetime}

Given a set a of timely ordered events \mbox{$X = \{x_t : t = 1, \cdots, n\}$}, where $x_t$ is the realization of $X$ at time $t$ for \mbox{$t \in T$}, we say that a timeseries is \textit{discrete} if $T$ are measurements taken at successive times spaced at uniform intervals \textit{w}, also referred to as sampling rate (defining the temporal granularity).

\begin{figure}[!ht]
        \centering
        \includegraphics[trim={0.5cm 0cm 0cm 0cm},width=0.47\textwidth]{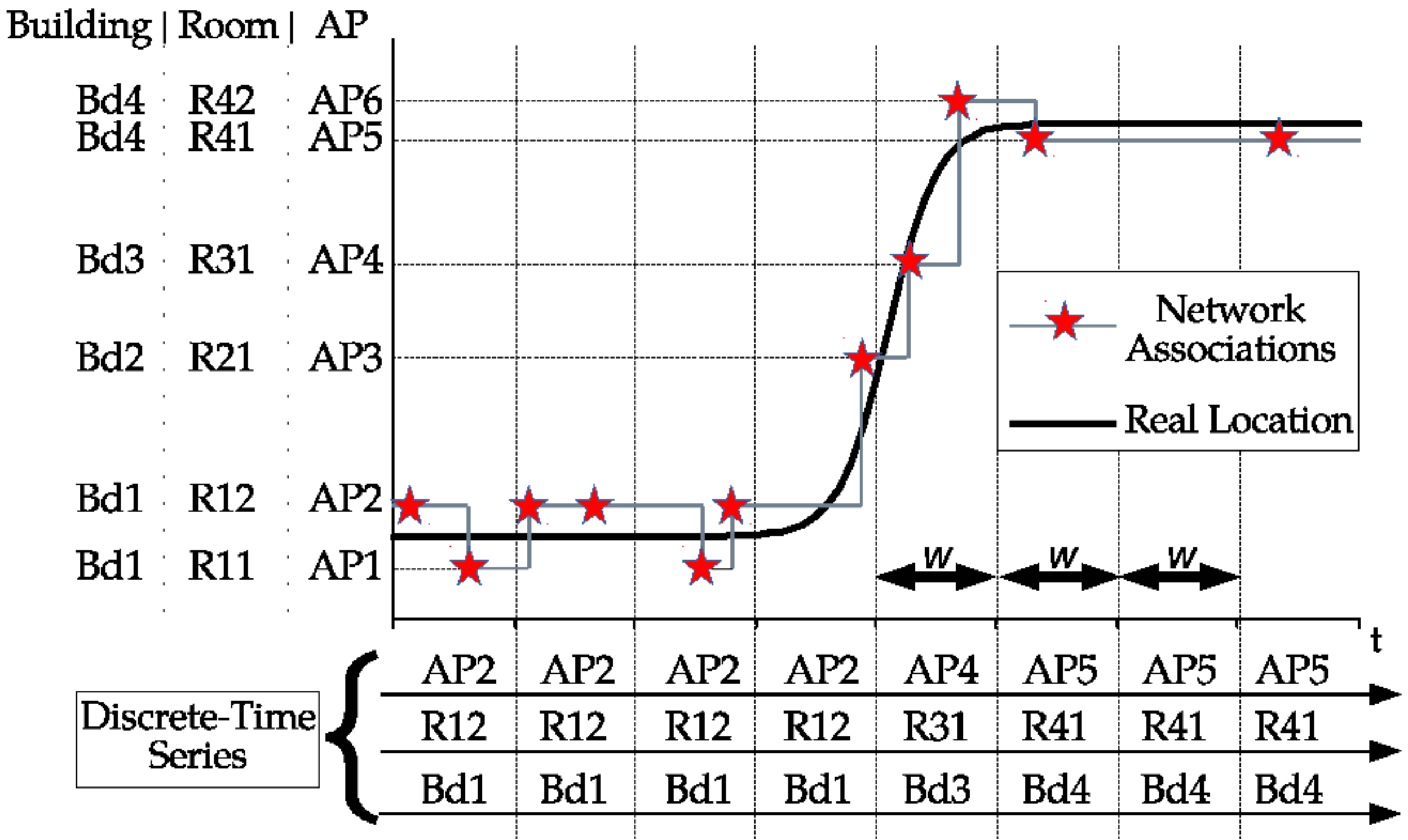}
        \caption{Location of the device is sampled at a constant rate.}
        \label{fig:discretetime}
\end{figure}

Figure \ref{fig:discretetime} depicts an example of how the real location of a device is sensed by the wireless management system through AP associations (red stars) and finally how the discrete-time series is obtained. For a given sampling time window \textit{w}, our \textit{discrete-time} series may result in different sequences depending on whether we choose an AP or a building as the level of spatial resolution.

From Figure \ref{fig:discretetime}, for the first 4 time steps the device switched its associated AP without a real location change. This switch in AP association can be triggered by the mobile device (e.g. stronger wireless signal) or by the network management system (e.g. load balancing).

Note that it is important to define the resolution for \textit{space} and \textit{time}, \textit{i.e.,} how big a location is in space (or point-of-interest) and how often we are going to sample from the input signal. In this example, larger values of \textit{w} could eliminate this \textit{ping-pong} effect of switching between APs without actually moving, but also cause loss of information when the user transits from one location to another. On the contrary, very small values of \textit{w} could over-sample long periods when the user is not moving. Similarly, different values of spatial resolution could mitigate noise but eliminate information from the traces. Choosing these parameters is often influenced by the characteristics of the available dataset as well as the targeted application of the study.

\subsubsection*{Step Value}
A weighing mechanism is used to pick the corresponding location to represent a time step. During a time interval, we weigh every observed location of the device with the duration of time at that location and pick the one with the highest weight to represent that step. We assign a user to a specific location $\ell$ in the time interval $\delta t$ between an association at $\ell$ and the next association at any other location, but only if $\delta t < t_{\text{max}}$. After $t_{\text{max}}$ the device will be in an \textit{unknown} state \cite{Song2010limits} until the next network event which will reveal its location for future steps.

\subsection{Experiments}
The design of our experiments is based on our study's questions: 
i. How different are \textit{Flutes} and \textit{Cellos} in terms of predictability? 
ii. How does the predictability of these device types change with different spatio-temporal granularity?
iii. Does the choice of method or predictor significantly alter the answers to aforementioned questions?
Thus, we evaluated a matrix, involving \textit{combinations} of the following dimensions:
\begin{itemize}
    \item Device Types: \textit{Flutes} vs. \textit{Cellos}.
    \item Temporal Resolutions: 5 min, 15 min, 30 min, 1 hour and 2 hours.
    \item Spatial Resolutions: Access Points, and Buildings.
    \item Methods: A. Well known sequence prediction algorithms from machine learning literature (Markov Chains, Neural Networks) B. Entropy-based Estimations of predictability upper-bounds.
\end{itemize}
The experiments were implemented in Python, the neural networks were implemented using Tensorflow \cite{tensorflow2015-whitepaper} and Keras.
Training is carried out in an \textit{online} manner and the evaluation is through providing a sliding window of $k$ observations to the predictor and testing the prediction correctness of the next symbol. The \textit{fraction of correct next symbol predictions} is the prediction accuracy metric.

\section{Experiment Results}
\label{sec:exp}

\subsection{Spatio-Temporal Resolutions}
To answer the first two questions of this study, particularly "ii. How does the predictability of these device types change with different \textit{spatio-temporal granularity}?",
Table \ref{tab:exp_spatio_temporal} summarizes the median accuracy of an LSTM predictor for Flutes and Cellos with different spatial and temporal granularity.

The choice of granularity is application-dependent, for example, to predict foot traffic at buildings and congestion planning based on density, building level analysis is more appropriate.
Cellos show more predictable behavior overall, as the fraction of correct next symbol predictions is higher for Cellos across the board.
At the AP level, with longer time bins, the accuracy for both Flutes and Cellos decreases. This observation is in line with previous findings \cite{Smith2014}. At 15min time intervals, the difference between Flutes and Cellos is at its maximum and drops and remains stable for longer time intervals.
At the building level, the accuracy follows a less regular pattern but both Flutes and Cellos are most predictable at 5min intervals (due to repeats of the same location in the sequence). Cellos' accuracy drops for 30min bins and goes back up again. On the other hand, Flutes are more predictable in 30min bins than 15min, 1h or 2h bins.

\begin{figure}[ht!]
	\centering
	\includegraphics[width=0.47\textwidth]{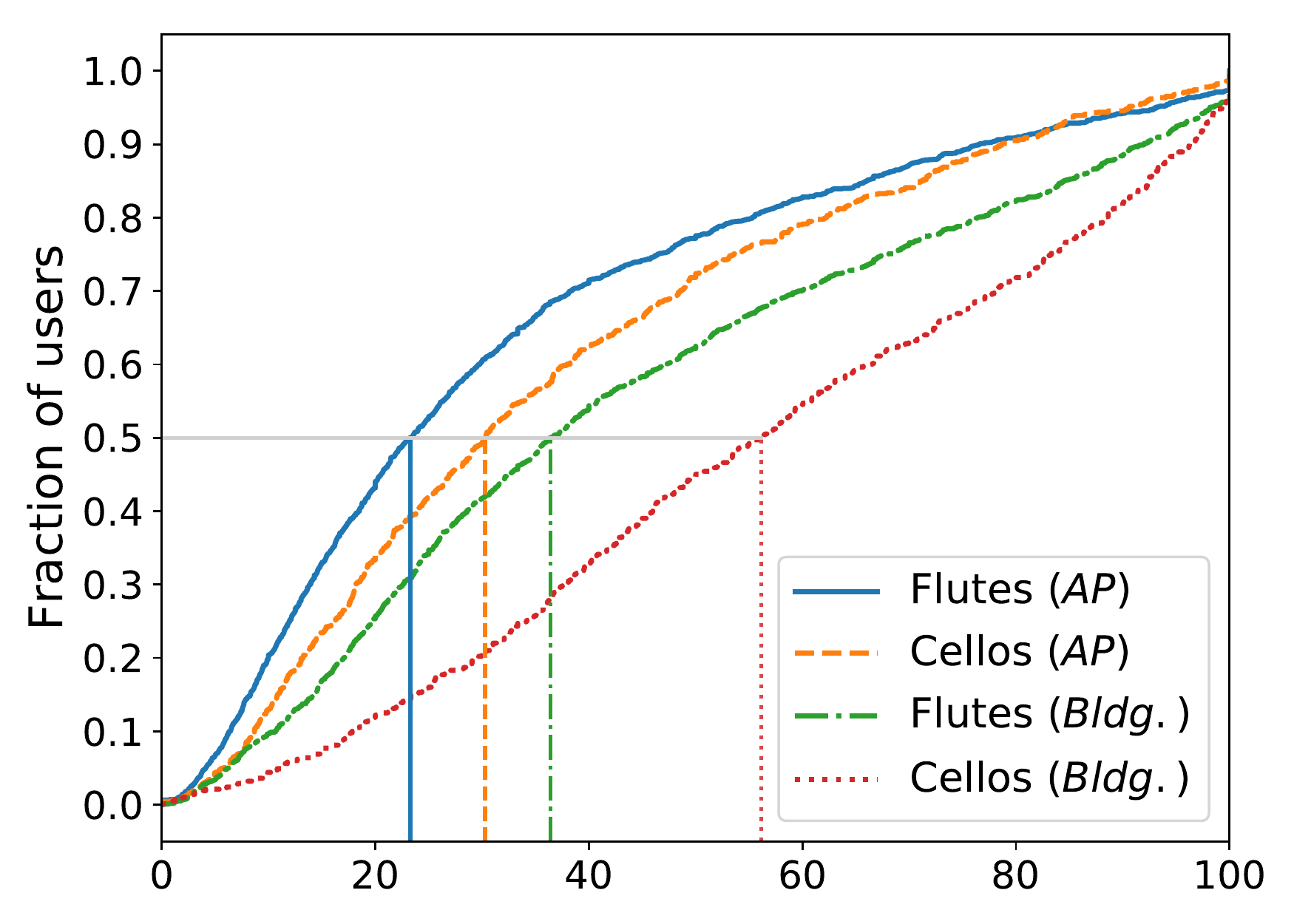}
	\caption{ECDF of LSTM Prediction Accuracy for \textit{Flutes} \& \textit{Cellos} at AP and Building spatial levels (all temporal levels combined, vertical lines denote medians, sequence length 40).}
	\label{fig:cdf_lstm_flutes_cellos}
\end{figure}

Looking across all temporal bins, Fig \ref{fig:cdf_lstm_flutes_cellos} presents the empirical cumulative distribution function (ECDF) of prediction accuracy at AP and building spatial granularity. The "sit-to-use" \textit{Cellos} show significantly higher predictability at every percentile; this is reasonable given their lower mobility \cite{Alipour2018} and mode of usage. In fact, prediction accuracy is highly correlated with other mobility and network traffic features of mobile wireless users, we will take a brief look at these correlations in Section \ref{sec:discussion} and Fig \ref{fig:corr}.

\begin{table}[!ht]
\centering
\caption{Median Accuracy of LSTM (sequence len. 40) for \textit{Flutes} vs \textit{Cellos}, 5min-2h temporal and AP/Bldg spatial granularity.}
\label{tab:exp_spatio_temporal}
\begin{tabular}{|c|c|c|c|c|}
\hline
       & \multicolumn{2}{c|}{AP} & \multicolumn{2}{c|}{Building} \\ \hline
       & F          & C          & F             & C             \\ \hline
5 min  & 33.22      & 42.25      & 44            & 63.4          \\ \hline
15 min & 21.42      & 36.9       & 34.53         & 58.06         \\ \hline
30 min & 21.88      & 27.39      & 39.56         & 50.78         \\ \hline
1 hour & 19.67      & 24.33      & 32.62         & 52.03         \\ \hline
2 hour & 17.17      & 22.5       & 32.6          & 59.62         \\ \hline
\end{tabular}
\end{table}

\begin{table*}[!ht]
\caption{Summary of Median Accuracy for \textit{Flutes} vs \textit{Cellos} with different methods (Diff is $Cellos - Flutes$) and sequence lengths for 15min and 1h time windows.}
\setlength\extrarowheight{2pt}
\label{tab:exp_cmp_methods}
\resizebox{\textwidth}{!}{%
\begin{tabular}{cc|c|c|c|c|c|c|c|c|c|c|c|c|}
\cline{3-14}
\multicolumn{1}{l}{}                        &           & \multicolumn{3}{c|}{AP, 1h}                                                                                  & \multicolumn{3}{c|}{Bldg., 1h}                                                                              & \multicolumn{3}{c|}{AP, 15min}                                                                              & \multicolumn{3}{c|}{Bldg., 15min}                                                                         \\ \hline
\multicolumn{1}{|l|}{Seq Len}               & Predictor & F                          & C                          & Diff                                               & F                          & C                         & Diff                                               & F                         & C                          & Diff                                               & F                         & C                         & Diff                                              \\ \Xhline{3\arrayrulewidth}
\multicolumn{1}{|c|}{}                      & MC        & 21.05                      & 25.95                      & \cellcolor[HTML]{B7F6B7}+4.90                      & 38.25                      & 53.50                     & \cellcolor[HTML]{B7F6B7}+15.25                     & 61.72                     & 70.30                      & \cellcolor[HTML]{B7F6B7}+8.58                      & 75.00                     & 87.60                     & \cellcolor[HTML]{B7F6B7}+12.60                    \\ \cline{2-14} 
\multicolumn{1}{|c|}{}                      & LSTM      & 21.62                      & 25.00                      & \cellcolor[HTML]{B7F6B7}+3.38                      & 35.03                      & 50.00                     & \cellcolor[HTML]{B7F6B7}+14.97                     & 40.00                     & 44.56                      & \cellcolor[HTML]{B7F6B7}+4.56                      & 52.44                     & 65.56                     & \cellcolor[HTML]{B7F6B7}+13.12                    \\ \cline{2-14} 
\multicolumn{1}{|c|}{\multirow{-3}{*}{5}}   & CNN       & 16.45                      & 24.27                      & \cellcolor[HTML]{B7F6B7}+7.82                      & 34.94                      & 50.00                     & \cellcolor[HTML]{B7F6B7}+15.06                     & 50.00                     & 59.80                      & \cellcolor[HTML]{B7F6B7}+9.80                      & 64.60                     & 76.94                     & \cellcolor[HTML]{B7F6B7}+12.34                    \\ \hline
\multicolumn{1}{|c|}{}                      & MC        & 17.98                      & 25.6                       & \cellcolor[HTML]{B7F6B7}+7.62                      & 36.72                      & 50.28                     & \cellcolor[HTML]{B7F6B7}+13.56                     & 52.25                     & 61.97                      & \cellcolor[HTML]{B7F6B7}+9.72                      & 68.00                     & 82.25                     & \cellcolor[HTML]{B7F6B7}+14.25                    \\ \cline{2-14} 
\multicolumn{1}{|c|}{}                      & LSTM      & 20.83                      & 26.31                      & \cellcolor[HTML]{B7F6B7}+5.48                      & 37.50                      & 50.66                     & \cellcolor[HTML]{B7F6B7}+13.16                     & 31.14                     & 44.62                      & \cellcolor[HTML]{B7F6B7}+13.48                     & 45.38                     & 64.56                     & \cellcolor[HTML]{B7F6B7}+19.18                    \\ \cline{2-14} 
\multicolumn{1}{|c|}{\multirow{-3}{*}{10}}  & CNN       & 18.06                      & 22.62                      & \cellcolor[HTML]{B7F6B7}+4.56                      & 36.20                      & 52.03                     & \cellcolor[HTML]{B7F6B7}+15.83                     & 49.20                     & 58.80                      & \cellcolor[HTML]{B7F6B7}+9.60                      & 64.56                     & 74.00                     & \cellcolor[HTML]{B7F6B7}+9.44                     \\ \hline
\multicolumn{1}{|c|}{}                      & MC        & 18.1                       & 24.52                      & \cellcolor[HTML]{B7F6B7}+6.42                      & 36.28                      & 49.94                     & \cellcolor[HTML]{B7F6B7}+13.66                     & 38.50                     & 48.22                      & \cellcolor[HTML]{B7F6B7}+9.72                      & 57.30                     & 74.94                     & \cellcolor[HTML]{B7F6B7}+17.64                    \\ \cline{2-14} 
\multicolumn{1}{|c|}{}                      & LSTM      & 21.22                      & 24.19                      & \cellcolor[HTML]{B7F6B7}+2.97                      & 36.12                      & 50.78                     & \cellcolor[HTML]{B7F6B7}+14.66                     & 29.17                     & 41.00                      & \cellcolor[HTML]{B7F6B7}+11.83                     & 43.62                     & 61.47                     & \cellcolor[HTML]{B7F6B7}+17.85                    \\ \cline{2-14} 
\multicolumn{1}{|c|}{\multirow{-3}{*}{20}}  & CNN       & 18.44                      & 23.60                      & \cellcolor[HTML]{B7F6B7}+5.16                      & 35.28                      & 50.00                     & \cellcolor[HTML]{B7F6B7}+14.72                     & 37.84                     & 48.12                      & \cellcolor[HTML]{B7F6B7}+10.28                     & 50.00                     & 65.00                     & \cellcolor[HTML]{B7F6B7}+15.00                    \\ \hline
\multicolumn{1}{|c|}{}                      & MC        & 17.88                      & 23.61                      & \cellcolor[HTML]{B7F6B7}+5.73                      & 35.1                       & 48.56                     & \cellcolor[HTML]{B7F6B7}+13.46                     & 27.97                     & 31,00                      & \cellcolor[HTML]{B7F6B7}+3.03                      & 47.12                     & 65.80                     & \cellcolor[HTML]{B7F6B7}+18.68                    \\ \cline{2-14} 
\multicolumn{1}{|c|}{}                      & LSTM      & 19.67                      & 24.33                      & \cellcolor[HTML]{B7F6B7}+4.66                      & 32.62                      & 52.03                     & \cellcolor[HTML]{B7F6B7}+19.41                     & 23.30                     & 39.40                      & \cellcolor[HTML]{B7F6B7}+16.10                     & 33.97                     & 59.03                     & \cellcolor[HTML]{B7F6B7}+25.06                    \\ \cline{2-14} 
\multicolumn{1}{|c|}{\multirow{-3}{*}{40}}  & CNN       & 18.75                      & 23.97                      & \cellcolor[HTML]{B7F6B7}+5.22                      & 35.25                      & 52.50                     & \cellcolor[HTML]{B7F6B7}+17.25                     & 27.62                     & 44.70                      & \cellcolor[HTML]{B7F6B7}+17.08                     & 41.25                     & 62.10                     & \cellcolor[HTML]{B7F6B7}+20.85                    \\ \Xhline{3\arrayrulewidth}
\multicolumn{1}{c|}{}                      & LZ        & \multicolumn{1}{l|}{46.90} & \multicolumn{1}{l|}{52.60} & \multicolumn{1}{l|}{\cellcolor[HTML]{B7F6B7}+5.70} & \multicolumn{1}{l|}{58.78} & \multicolumn{1}{l|}{66.40} & \multicolumn{1}{l|}{\cellcolor[HTML]{B7F6B7}+7.62} & \multicolumn{1}{l|}{72.70} & \multicolumn{1}{l|}{76.06} & \multicolumn{1}{l|}{\cellcolor[HTML]{B7F6B7}+3.36} & \multicolumn{1}{l|}{79.60} & \multicolumn{1}{l|}{79.10} & \multicolumn{1}{l|}{\cellcolor[HTML]{FBE3E2}-0.50} \\ \cline{2-14} 
\multicolumn{1}{c|}{\multirow{-2}{*}{}}    & BWT       & \multicolumn{1}{l|}{66.44} & \multicolumn{1}{l|}{69.44} & \multicolumn{1}{l|}{\cellcolor[HTML]{B7F6B7}+3.00} & \multicolumn{1}{l|}{73.70} & \multicolumn{1}{l|}{79.90} & \multicolumn{1}{l|}{\cellcolor[HTML]{B7F6B7}+6.20} & \multicolumn{1}{l|}{83.30} & \multicolumn{1}{l|}{88.06} & \multicolumn{1}{l|}{\cellcolor[HTML]{B7F6B7}+4.76} & \multicolumn{1}{l|}{88.60} & \multicolumn{1}{l|}{92.20} & \multicolumn{1}{l|}{\cellcolor[HTML]{B7F6B7}+3.60} \\ \cline{2-14} 

\end{tabular}%
}
\end{table*}

\subsection{Comparison of Methods}

To answer the third question of this study, "iii. Does the choice of method or predictor significantly alter the answers to aforementioned questions?", 
here we compare the experiment results for different methods: 
1) \textit{MC}: Markov Chain 
2) \textit{LSTM}: A type of recurrent neural network 
3) \textit{CNN}: 1D Convolutional Neural Network 
4) \textit{Hr\_LZ}: Theoretical predictability based on the Lempel-Ziv (LZ) entropy estimator 
5) \textit{Hr\_BWT}: Theoretical predictability based on the Burrows-Wheeler transform (BWT) entropy estimator.
A summary of comparisons is presented in Table \ref{tab:exp_cmp_methods}, for temporal granularity of 1h and 15min, highlighting the difference of \textit{Cellos} - \textit{Flutes}.

In all cases \textit{Cellos} are more predictable than \textit{Flutes}, regardless of the choice of method (with a minor exception of LZ predictor at 15min time and building level which might be due to intrinsic instability of LZ based estimator). 
The difference in median accuracy for \textit{Flutes} vs \textit{Cellos} is up to 25\% (Building level, 15min window, sequence length 40, \textit{Flutes} 33.97\% vs \textit{Cellos} 59.03\%).
Other temporal choices result in a similar pattern. 
Another notable observation is that while the neural networks are more complex, and require vastly more computing power, they only achieve modest increase compared to Markov Chains in \textit{some} scenarios (e.g., \textit{Cellos}, at the Bldg. level and Seq. Len. 40, from 48.56\% to 52.5\%). This is a trade-off that needs to be considered in the design of predictive caching systems.
In addition, increasing the sequence length $k$ (i.e. the number of previous time steps available to the predictor) impacts the Markov Chain model more than the neural networks. This is particularly pronounced for 15min time window, in fact, the neural networks do not lose much accuracy from increasing sequence length 5 to 40 in case of the 1h time window.
Also, the theoretical LZ and BWT based estimators, show higher upper bounds compared with the best of the algorithms, with Seq. Len. 5 Markov Chains and CNNs being the closest practical algorithms for the 15min case. The predictors are far behind in the 1h case, suggesting room for improvement via tuning for specific time and space granularities.
The run time of LSTM is the longest, followed by CNN (not shown for brevity).


\section{Discussion \& Future Work}
\label{sec:discussion}

In this paper, we define our research problem as predicting the next symbol in a discrete-time series for users with two categories of devices. The accuracy is evaluated as the fraction of the next symbols predicted correctly. 

While some earlier studies investigated a similar problem setup, our study has notable implications. For example, across device types, predictability can vary significantly. Also, with larger time windows such as 1 hour, it is easy to miss short stays (since one location visit with a duration of 31 minutes would result in other locations in that 1h window being ignored). On the other hand, a short time window results in multiple repetitions of the same location in the sequence, potentially achieving high prediction accuracy even when the method is not predicting the \textit{transitions} well. 

It is important to consider the device type, context, and application in order to choose an appropriate time and space granularity; the best performing method differs across these dimensions.
Besides, the measured accuracy only considers an exact match to be correct, so even if the method predicts a nearby location to the actual location, it would count as incorrect. We plan to investigate measuring how far a predicted location is from the actual location and embed that information in the loss function of our neural networks for possible improvements in prediction.


\subsubsection*{Correlations with Mobility and Network Traffic}
\begin{figure}[ht!]
	\centering
	\includegraphics[width=0.47\textwidth]{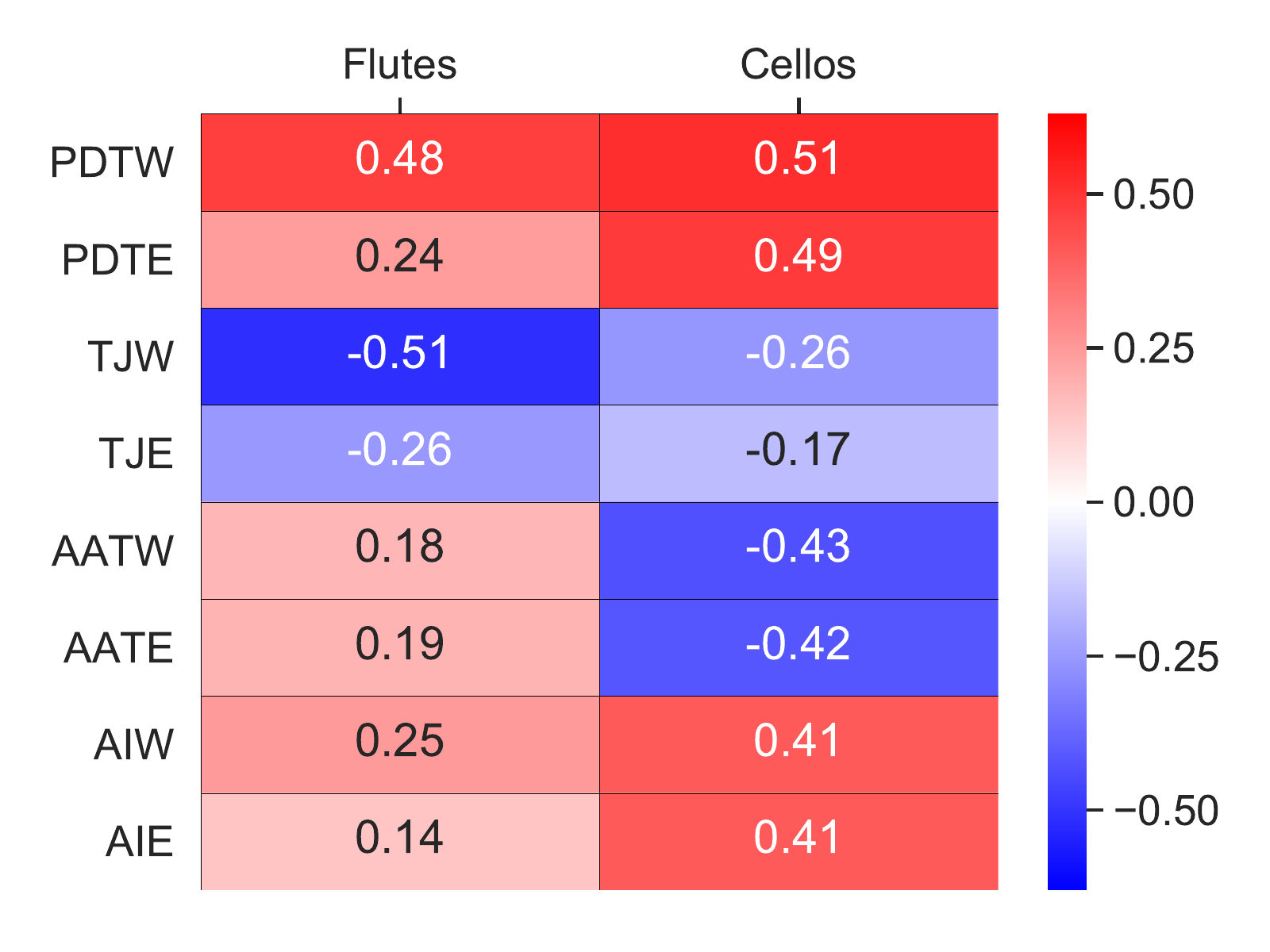}
	\caption{Pearson Correlation of Prediction Accuracy with several Mobility and Network Traffic Features.}
	\label{fig:corr}
\end{figure}

Figure \ref{fig:corr} shows the correlation of prediction accuracy with a sample of features that describe the mobility or network traffic of users. PDT(W/E) and TJ(W/E) are mobility features while AAT(W/E) and AI(W/E) are traffic features.
PDTW is the time spent at the user's preferred building (most common) on weekdays (PDTE for weekends). TJW is the total sum of jumps (distance) for the weekdays while TJE describes the same feature for weekends. AATW is the average of active time (as indicated by network usage) of the user for weekdays (AATE for weekends). AIW stands for average inter-arrival time of flows on weekdays, and AIE for weekends (\cite{Alipour2018, alipour2018learning}).

The results present significant correlations between the prediction accuracy, with not only the \textit{mobility features}, but also \textit{network traffic features}.
These correlations vary across device types (\textit{Flutes} vs \textit{Cellos}), and in time (\textit{Weekdays} vs \textit{Weekends}).
This is a very important observation for the design of \textit{predictive caching} systems, importantly, it might be possible to improve prediction of \textit{where} the user is going based on network traffic profile while noting the different modes of usage based on device types. We leave the investigation of such improvements to future work.

\subsubsection*{Integrated Mobility-Traffic Modeling}
Given the observed correlations, we hypothesize that use of \textit{predictability} as a feature in an integrated mobility-traffic generative model could lead to more realistic synthetic traces. 
Such a data-driven generative model would be an essential tool for network simulations and capacity planning. Notably, it can also be made \textit{privacy preserving}, since collected traces would be replaced with realistic synthetic data that captures mobility, network traffic, predictability, and their relationships.
Further study is beyond the scope of this work and is left for the future.


\section{Conclusion}
\label{sec:conclusion}
In this work, we sought to answer three questions:
i. How different are \textit{Flutes} and \textit{Cellos} in terms of predictability? 
ii. How does the predictability of these device types change with different \textit{spatio-temporal granularity}?
iii. Does the \textit{choice of method} or predictor significantly alter the answers to aforementioned questions?
For this purpose, we processed a large-scale dataset from a campus environment, and grouped the devices into two categories; and we chose a set of methods to make the comparisons, including Entropy-based estimators and popular algorithms such as Markov Chains and Neural Networks.

The results of experiments show the movements of \textit{Cellos} ("sit-to-use") are significantly more predictable than \textit{Flutes} (up to 25\% difference in accuracy).
This pattern is consistent across various temporal granularities (5 min to 2 hours), spatial granularities (Access Point and Building level), and for different methods (Markov Chains, Neural Networks, Entropy-based Estimators).
We illustrate that the performance of predictors depends strongly on the span of temporal bins. Markov Chains tend to outperform deep learning models in shorter time-bins while LSTMs and CNNs usually show a higher accuracy in longer time-bins. CNNs have mostly similar accuracy to LSTMs in the latter case but have significantly better run time on a modern GPU.
We also found significant correlations among prediction accuracy, \textit{mobility features}, and also \textit{network traffic features}, an important observation for the design of \textit{predictive caching} systems where it might be possible to improve mobility prediction based on network traffic profile.
We plan to further investigate the use of \textit{predictability as a feature} in an integrated mobility-traffic generative model, and its application in state-of-the-art predictive caching systems.

\section*{Acknowledgement}
This work was partially funded by NSF 1320694.
We gratefully acknowledge the support of NVIDIA Corp. with the donation of the Titan Xp GPU used for this research.

\nocite{*} 

\bibliographystyle{IEEEtran}
\bibliography{main}

\begin{thebibliography}{10}
\providecommand{\url}[1]{#1}
\csname url@samestyle\endcsname
\providecommand{\newblock}{\relax}
\providecommand{\bibinfo}[2]{#2}
\providecommand{\BIBentrySTDinterwordspacing}{\spaceskip=0pt\relax}
\providecommand{\BIBentryALTinterwordstretchfactor}{4}
\providecommand{\BIBentryALTinterwordspacing}{\spaceskip=\fontdimen2\font plus
\BIBentryALTinterwordstretchfactor\fontdimen3\font minus
  \fontdimen4\font\relax}
\providecommand{\BIBforeignlanguage}[2]{{%
\expandafter\ifx\csname l@#1\endcsname\relax
\typeout{** WARNING: IEEEtran.bst: No hyphenation pattern has been}%
\typeout{** loaded for the language `#1'. Using the pattern for}%
\typeout{** the default language instead.}%
\else
\language=\csname l@#1\endcsname
\fi
#2}}
\providecommand{\BIBdecl}{\relax}
\BIBdecl

\bibitem{Siris2016}
V.~Siris, X.~Vasilakos, and D.~Dimopoulos, ``Exploiting mobility prediction for
  mobility, popularity caching and dash adaptation,'' in \emph{WoWMoM}, 2016.

\bibitem{Lathia2015anatomy}
N.~Lathia, ``The anatomy of mobile location-based recommender systems,'' in
  \emph{Recommender Systems Handbook}.\hskip 1em plus 0.5em minus 0.4em\relax
  Springer, 2015.

\bibitem{Song2010limits}
C.~Song, Z.~Qu, N.~Blumm, and A.-L. Barab{\'a}si, ``Limits of predictability in
  human mobility,'' \emph{Science}, 2010.

\bibitem{Smith2014}
G.~Smith, R.~Wieser, J.~Goulding, and D.~Barrack, ``{A refined limit on the
  predictability of human mobility},'' \emph{PerCom}, 2014.

\bibitem{Cao2017infocom}
P.~Cao, G.~Li, A.~Champion, D.~Xuan, S.~Romig, and W.~Zhao, ``On human mobility
  predictability via {WLAN} logs,'' in \emph{Proc. INFOCOM}, Apr. 2017.

\bibitem{Li2014its}
Y.~Li, D.~Jin, P.~Hui, Z.~Wang, and S.~Chen, ``Limits of predictability for
  large-scale urban vehicular mobility,'' \emph{IEEE Transactions on
  Intelligent Transportation Systems}, 2014.

\bibitem{Wang2015}
J.~Wang, Y.~Mao, J.~Li, Z.~Xiong, and W.~X. Wang, ``{Predictability of road
  traffic and congestion in urban areas},'' \emph{PLoS ONE}, 2015.

\bibitem{Gallotti2013}
R.~Gallotti, A.~Bazzani, M.~D. Esposti, and S.~Rambaldi, ``{Entropic measures
  of individual mobility patterns},'' \emph{Journal of Statistical Mechanics:
  Theory and Experiment}, no.~10, 2013.

\bibitem{takaguchi2011predictability}
T.~Takaguchi, M.~Nakamura, N.~Sato, K.~Yano, and N.~Masuda, ``Predictability of
  conversation partners,'' \emph{Physical Review X}, 2011.

\bibitem{sinatra2014entropy}
R.~Sinatra and M.~Szell, ``Entropy and the predictability of online life,''
  \emph{Entropy}, vol.~16, no.~1, pp. 543--556, 2014.

\bibitem{Hanel2011}
R.~Hanel and S.~Thurner, ``{A comprehensive classification of complex
  statistical systems and an axiomatic derivation of their entropy and
  distribution functions},'' \emph{Epl}, vol.~93, no.~2, 2011.

\bibitem{Zhou2012}
X.~Zhou, Z.~Zhao, R.~Li, Y.~Zhou, and H.~Zhang, ``{The predictability of
  cellular networks traffic},'' in \emph{ISCIT 2012}, 2012.

\bibitem{Goulet-Langlois2017}
G.~Goulet-Langlois, H.~N.~Koutsopoulos, Z.~Zhao, and J.~Zhao, ``Measuring
  regularity of individual travel patterns,'' \emph{IEEE Transactions on
  Intelligent Transportation Systems}, 2017.

\bibitem{maier2010first}
G.~Maier, F.~Schneider, and A.~Feldmann, ``A first look at mobile hand-held
  device traffic,'' in \emph{PAM}.\hskip 1em plus 0.5em minus 0.4em\relax
  Springer, 2010.

\bibitem{chen2012network}
X.~Chen, R.~Jin, K.~Suh, B.~Wang, and W.~Wei, ``Network performance of smart
  mobile handhelds in a university campus wifi network,'' \emph{ACM IMC}, 2012.

\bibitem{Kumar2013}
U.~Kumar, J.~Kim, and A.~Helmy, ``{Changing patterns of mobile network (WLAN)
  usage: Smart-phones vs. laptops},'' \emph{IWCMC}, 2013.

\bibitem{Alipour2018}
B.~Alipour, L.~Tonetto, A.~Yi~Ding, R.~Ketabi, J.~Ott, and A.~Helmy, ``Flutes
  vs. cellos: Analyzing mobility-traffic correlations in large wlan traces,''
  in \emph{IEEE INFOCOM}, 2018.

\bibitem{Gao2008}
Y.~Gao, I.~Kontoyiannis, and E.~Bienenstock, ``{Estimating the entropy of
  binary time series: Methodology, some theory and a simulation study},''
  \emph{Entropy}, vol.~10, no.~2, pp. 71--99, 2008.

\bibitem{Cai2004}
H.~Cai, S.~R. Kulkarni, and S.~Verd{\'{u}}, ``{Universal entropy estimation via
  block sorting},'' pp. 1551--1561, 2004.

\bibitem{Song2004infocom-predictors}
L.~Song, D.~Kotz, R.~Jain, and X.~He, ``Evaluating location predictors with
  extensive {Wi-Fi} mobility data,'' in \emph{INFOCOM}, 2004.

\bibitem{lu2013approaching}
X.~Lu, E.~Wetter, N.~Bharti, A.~J. Tatem, and L.~Bengtsson, ``Approaching the
  limit of predictability in human mobility,'' \emph{Scientific reports},
  vol.~3, 2013.

\bibitem{lstm}
S.~Hochreiter and J.~Schmidhuber, ``Long short-term memory,'' \emph{Neural
  computation}, vol.~9, no.~8, pp. 1735--1780, 1997.

\bibitem{gru}
K.~Cho, B.~Van~Merri{\"e}nboer, C.~Gulcehre, D.~Bahdanau, F.~Bougares,
  H.~Schwenk, and Y.~Bengio, ``Learning phrase representations using rnn
  encoder-decoder for statistical machine translation,'' \emph{arXiv preprint
  arXiv:1406.1078}, 2014.

\bibitem{DLsurvey}
J.~Schmidhuber, ``Deep learning in neural networks: An overview,'' \emph{Neural
  networks}, vol.~61, pp. 85--117, 2015.

\bibitem{karatzoglou2018seq2seq}
A.~Karatzoglou, A.~Jablonski, and M.~Beigl, ``A seq2seq learning approach for
  modeling semantic trajectories and predicting the next location,'' in
  \emph{ACM SIGSPATIAL}, 2018.

\bibitem{tensorflow2015-whitepaper}
``{TensorFlow}: Large-scale machine learning on heterogeneous systems,''
  software available from tensorflow.org.

\bibitem{alipour2018learning}
B.~Alipour, M.~Al~Qathrady, and A.~Helmy, ``Learning the relation between
  mobile encounters and web traffic patterns: A data-driven study,'' in
  \emph{ACM MSWIM}, 2018.

\end{thebibliography}

\end{document}